\documentclass[aps,prl,twocolumn,superscriptaddress,floatfix]{revtex4-2}

\usepackage{amsmath,amssymb,bm}
\usepackage{graphicx}
\usepackage[dvipsnames]{xcolor}
\usepackage{dcolumn}
\usepackage{multirow}
\usepackage[colorlinks=true,linkcolor=NavyBlue,citecolor=NavyBlue,
            urlcolor=NavyBlue]{hyperref}

\graphicspath{{./}{figures/}{figures/}{figures/figures/}}

\begin{document}

\title{Zeeman Quantum Geometry of the Layer Pseudo-spin}

\author{Bingxuan Jiang}
\affiliation{School of Physics and Electronics, Hunan University, Changsha 410082, China}
\author{Shihao Zhang}
\thanks{Corresponding author: zhangshh@hnu.edu.cn}
\affiliation{School of Physics and Electronics, Hunan University, Changsha 410082, China}
\date{\today}

\begin{abstract}
Two-dimensional layers coupled by position-dependent interlayer tunnelling host a layer pseudo-spin whose polarization is a controllable internal degree of freedom. We introduce the quantum geometry of this layer pseudo-spin: a mixed tensor formed from the interband Berry connection and the layer matrix element, which governs the response to an oscillating in-plane electric field. Because the layer pseudo-spin is an Ising-type scalar invariant under threefold rotation, its selection rules differ from those of a pseudovector spin. In a minimal two-band model, the layer-pseudo-spin quantum metric and Zeeman--Berry curvature are nonzero, while their zone integrals vanish. An in-plane drive therefore leaves the spatially averaged polarization unchanged at linear order but imprints a moir\'{e}-periodic intracell pattern, directly imageable by layer-resolved tunnelling. For twisted MoTe$_2$ bilayer, we quantify the local dipole modulation and identify the rectified, helicity-odd, and third-harmonic channels through which a net response appears at higher order.
\end{abstract}
\maketitle

\emph{Introduction.} Two-dimensional layers that are stacked with a relative twist or shift are coupled by an interlayer tunnelling whose amplitude varies with position.  This position-dependent hybridization ties the layer index to the superlattice and turns the layer pseudo-spin $\boldsymbol{\tau}=(\tau_{x},\tau_{y},\tau_{z})$ into a quantum number of the low-energy states: $\tau_{z}$ measures the difference of the weights of the two layers in a Bloch state, $\tau_{x,y}$ describe the interlayer coherence, and the superlattice acts on them as a pseudo-magnetic field \cite{Wu2019}.  An out-of-plane displacement field $V_{z}$ makes the two layers inequivalent, and the layer polarization becomes a controllable internal degree of freedom: the order parameter of interfacial and moir\'{e} ferroelectricity \cite{Weston2022} and the out-of-plane structure on which the correlated states of the flat bands are built \cite{Park2023,bistritzer2011moire,cao2018correlated,cao2018unconventional,balents2020superconductivity,song2022magic,liu2019pseudo,moon2012energy,zhang2022correlated,koshino2018maximally,liu2019quantum,kang2019strong,zhang2022spin,kang2018symmetry,tarnopolsky2019origin,bultinck2020ground}.  A uniform out-of-plane field polarizes the two layers, and the natural question is how this degree of freedom responds to a uniform in-plane field, which acts on the two layers in the same way and couples to the interlayer coherence $\tau_{x,y}$ rather than to the layer populations $\tau_{z}$.

Quantum geometry associates with the Bloch states a metric and a curvature, the real and imaginary parts of the quantum geometric tensor \cite{ProvostVallee,Berry,Xiao2010}, which control a broad range of linear and nonlinear responses \cite{Sodemann2015,Morimoto2016,Peotta2015,Torma2022}.  For a layer degree of freedom the relevant object is a mixed tensor, carrying one momentum derivative and one layer derivative, whose form follows from a single difference with respect to the spin case: a spin is a pseudovector that rotates with the lattice, whereas the layer polarization is a scalar that is invariant under the threefold rotation of the superlattice.  A spin polarized by an in-plane field acquires a net moment, while the layer degree of freedom responds as a spatial pattern, with the layer character modulated inside each unit cell, and it is this modulation that a local probe of one layer measures.

In this work, we show that an oscillating in-plane field controls the layer degree of freedom, and that its response is governed by the quantum geometry of the layer pseudo-spin. The minimal two-band layer-pseudo-spin model shows that the mixed tensor built from the interband Berry connection and the layer matrix element is nonzero at every $\mathbf{k}$, and its imaginary part, the layer-pseudo-spin Zeeman--Berry curvature, is the momentum-space gradient of the layer polarization. The drive converts this momentum-space geometry into a real-space pattern: it leaves the spatially averaged layer polarization untouched at linear order in the field and imprints an intracell pattern instead, so that a layer-resolved probe, such as a scanning tunnelling measurement of one layer, images the pattern directly. Furthermore, a spatially averaged response appears at higher order in the field, and we identify every channel in which it does: a rectified response linear in $V_{z}$, a helicity-odd response, also linear in $V_{z}$, that is opposite in the two valleys, and a response at the third harmonic of the drive with a sixfold dependence on the polarization angle.  We evaluate the pattern for a twisted transition-metal dichalcogenide homobilayer, where it is largest in the window in which the moir\'{e} bands are flattest.

\emph{Minimal two-band model and Zeeman quantum geometry.}  We begin our discussions by general two-band Hamiltonian for bilayer system
\begin{equation}
\begin{split}
H_{\min}(\mathbf{k})=&\,\varepsilon(\mathbf{k})\,\tau_{0}
+\Delta_{z}(\mathbf{k})\,\tau_{z}\\
&+\Delta_{\parallel}(\mathbf{k})\bigl[\cos\gamma(\mathbf{k})\,\tau_{x}
+\sin\gamma(\mathbf{k})\,\tau_{y}\bigr],
\end{split}
\label{eq:minimal}
\end{equation}
where $\Delta_{\parallel}\ge0$ and $\gamma$ are the magnitude and the phase of the in-plane component of the layer pseudo-field.  Diagonalising Eq.~(\ref{eq:minimal}) gives $\langle\tau_{z}\rangle_{\pm}=\pm\cos\theta$ with $\tan\theta=\Delta_{\parallel}/\Delta_{z}$ and $\langle\tau_{x,y}\rangle_{\pm}=\pm\sin\theta\,(\cos\gamma,\sin\gamma)$: the band carries a layer-pseudo-spin texture whose polar and azimuthal angles are $\theta$ and $\gamma$. Then the symmetry of the superlattice fixes the form of $\gamma$.  Under $C_{3z}$ rotation operator, the layer pseudo-spin operators stay invariant, since a rotation about $\hat{z}$ neither interchanges the layers nor touches the layer character, while momentum rotates, so that the pseudo-field components are invariant functions of $\mathbf{k}$.  The leading in-plane term allowed by the threefold symmetry is the third harmonic, $\Delta_{\parallel}e^{i\gamma}=\lambda\,(k_{x}+ik_{y})^{3}/k_{0}^{3},
\gamma=3\varphi+\gamma_{0},
\label{eq:minimalfield}
$
with $\varphi$ the azimuthal angle of $\mathbf{k}$, $k_{0}$ the momentum scale below which Eq.~(\ref{eq:minimal}) is the low-energy description of the band, $\lambda$ the amplitude of the in-plane pseudo-field and $\gamma_{0}$ a constant phase.  
The winding of the pseudo-field is what gives the layer degree of freedom a nontrivial geometry.

The linear response of the layer pseudo-spin is governed by the Zeeman quantum geometry tensor
\begin{equation}
z^{\mu\nu}_{nm}=a^{\nu}_{nm}s^{\mu}_{mn},
\label{eq:ztensor}
\end{equation}
built from the interband Berry connection and the layer-pseudo-spin matrix element,
\begin{equation}
a^{\nu}_{nm}=i\langle u_{n}|\partial_{k_{\nu}}u_{m}\rangle,\qquad
s^{\mu}_{mn}=\langle u_{m}|\tau_{\mu}|u_{n}\rangle,
\end{equation}
in which the first index labels the observable and the second the field direction, so that $S^{\mu}=A^{\mu\nu}E^{\nu}$.  The real and imaginary parts of $z^{\mu\nu}_{nm}$ define the Zeeman quantum metric and the Zeeman--Berry curvature (Supplemental Material, Eq.~(\ref{eq:SM:Zdef})).  Following Ref.~\cite{ZeemanOrigin}, $z$ is the mixed term of the quantum distance on a Hilbert space parametrized by both the crystal momentum and the layer degree of freedom (Supplemental Material, Sec.~\ref{sec:SM:geometry} ).  Solving the quantum Liouville equation to first order in the field (Supplemental Material, Sec.~\ref{sec:SM:kubo} ) gives the Kubo formula
\begin{equation}
\begin{split}
\chi^{\mu\nu}(\omega)=\int_{\mathrm{mBZ}}\!d^{2}k
\sum_{n>m}f_{nm}\Bigl[&\frac{z^{\mu\nu}_{nm}}{\omega-\varepsilon_{mn}+i\eta}\\
&-\frac{z^{*\,\mu\nu}_{nm}}{\omega+\varepsilon_{mn}+i\eta}\Bigr],
\end{split}
\label{eq:kubo}
\end{equation}
with $S^{\mu}(t)=\operatorname{Re}\,[\chi^{\mu\nu}E^{\nu}e^{-i\omega t}]$, $f_{nm}=f_{n}-f_{m}$ and $\varepsilon_{mn}=\varepsilon_{n}-\varepsilon_{m}>0$. For $\hbar\omega$ well below the interband spacing, Eq.~(\ref{eq:kubo}) separates into an in-phase (quantum-metric) and a quadrature (Zeeman--Berry) part,
\begin{align}
A^{\mu\nu}_{\cos}&=
-\int_{\mathrm{mBZ}}\!d^{2}k\sum_{n>m}f_{nm}
\frac{2\,\mathcal{Q}^{\mu\nu}_{nm}}{\varepsilon_{mn}},
\label{eq:Acos}\\
A^{\mu\nu}_{\sin}&=
-\int_{\mathrm{mBZ}}\!d^{2}k\sum_{n>m}f_{nm}
\frac{\mathcal{Z}^{\mu\nu}_{nm}\,\omega}{\varepsilon_{mn}^{2}},
\label{eq:Asin}
\end{align}
so that $\chi^{\mu\nu}\simeq A^{\mu\nu}_{\cos}+iA^{\mu\nu}_{\sin}$, with the in-phase part $\omega$-independent and the quadrature part linear in $\omega$.

For the model of Eq.~(\ref{eq:minimal}) the tensor follows in closed form (Supplemental Material, Sec.~\ref{sec:SM:minimal} ),
\begin{equation}
z^{z\nu}_{+-}=\tfrac{1}{2}\sin^{2}\!\theta\,
\partial_{k_{\nu}}\gamma+\frac{i}{2}\sin\theta\,\partial_{k_{\nu}}\theta,
\label{eq:minimalz}
\end{equation}
so that the layer-pseudo-spin quantum metric and the Zeeman--Berry curvature are
\begin{equation}
\mathcal{Q}^{z\nu}=\tfrac{1}{2}\sin^{2}\!\theta\,\partial_{k_{\nu}}\gamma,
\qquad
\mathcal{Z}^{z\nu}=\sin\theta\,\partial_{k_{\nu}}\theta
=-\partial_{k_{\nu}}\langle\tau_{z}\rangle_{+}.
\label{eq:minimalQZ}
\end{equation}
We note that both $\mathcal{Q}^{z\nu}$ and $\mathcal{Z}^{z\nu}$ are nonzero at every $\mathbf{k}$.
The minimal model also displays the character of the response: the drive modulates the layer character at fixed zone-averaged $\langle\tau_{z}\rangle$, coupling to the $\mathbf{k}$-resolved geometry of Eq.~(\ref{eq:minimalQZ}), whose amplitude is anisotropic in the zone and follows the third harmonic of $\varphi$.  Inside a moir\'{e} superlattice the pseudo-field is itself periodic, and the same geometry appears in real space as a pattern inside each moir\'{e} unit cell; that is the system we study next.

\emph{Layer-pseudo-spin response in the twisted MoTe$_2$ homobilayer.}  Following Ref.~\cite{Wu2019} we describe the valence states of a twisted MoTe$_2$ homobilayer by a two-band Hamiltonian in the layer pseudo-spin basis $\boldsymbol{\tau}=(\tau_{x},\tau_{y},\tau_{z})$, with $\tau_{z}=+1$ ($-1$) for the bottom (top) layer.  Ising spin--orbit coupling locks the spin to the valley, so that we keep a single band per valley and work with a spinless continuum model: the spin is frozen, it enters only through the valley-dependent parameters, and no spin--orbit term appears explicitly. The effective Hamiltonian in the $+K$ valley is described by
\begin{equation}
\mathcal{H}_{+}(\mathbf{r},\mathbf{k})=
-\frac{\hbar^{2}\mathbf{k}^{2}}{2m^{*}}+\Delta_{+}(\mathbf{r})
+\boldsymbol{\Delta}(\mathbf{r})\cdot\boldsymbol{\tau}
+\frac{V_{z}}{2}\tau_{z}
\label{eq:model}
\end{equation}
up to the layer-dependent momentum shifts that fold the two Dirac points onto the moir\'{e} Brillouin zone (mBZ), with $\Delta_{+}$ the layer-symmetric part of the moir\'{e} potential and
\begin{equation}
\boldsymbol{\Delta}(\mathbf{r})=\left(\operatorname{Re}\Delta_{T}^{\dagger},\;
\operatorname{Im}\Delta_{T}^{\dagger},\;
\frac{\Delta_{\mathfrak{b}}-\Delta_{\mathfrak{t}}}{2}\right)
\label{eq:pseudofield}
\end{equation}
the layer pseudo-magnetic field (Supplemental Material, Eqs.~(\ref{eq:SM:potentials})--(\ref{eq:SM:moirelength}) ); the $-K$ valley follows by time reversal.  Because the moir\'{e} phase is $\psi\simeq-90^{\circ}$, the out-of-plane component $\Delta_{z}\propto\sin\psi$ dominates the in-plane one $\propto\cos\psi$: the layer pseudo-spin precesses in a nearly out-of-plane pseudo-magnetic field and the flat bands are almost fully layer polarized.

Now we drive the system with an oscillating in-plane electric field of frequency $\omega$, $\mathbf{E}(t)=-\partial\mathbf{A}/\partial t$, introduced through the minimal coupling $\mathbf{k}\to\mathbf{k}-\mathbf{A}(t)$ ($e=\hbar=1$), taking
\begin{equation}
\begin{split}
\mathbf{A}^{\mathrm{lin}}(t)&=A_{0}\sin(\omega t)\,\hat{x},\\
\mathbf{A}^{\mathrm{circ}}(t)&=A_{0}\bigl(\cos\omega t\,\hat{x}
+\sin\omega t\,\hat{y}\bigr),
\end{split}
\label{eq:drive}
\end{equation}
so that the field amplitude is $E_{0}=\omega A_{0}$.  We use weak, THz-scale fields ($E_{0}$ of a few kV\,cm$^{-1}$), well inside the linear-response regime of the band, and take the field to be strictly in plane, so that $V_{z}$ stays fixed.  The two drives differ in symmetry: $\mathbf{A}^{\mathrm{lin}}$ is even under time reversal, whereas $\mathbf{A}^{\mathrm{circ}}$ is mapped onto the opposite helicity. This is the origin of the two distinct selection rules derived below.

\begin{figure}[tbp]
\centering
\includegraphics[width=\columnwidth]{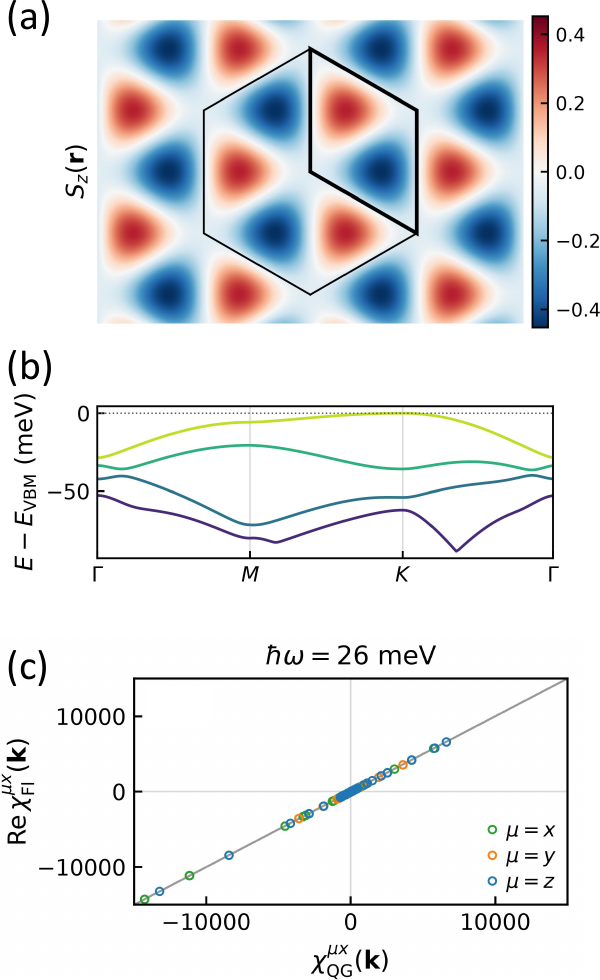}
\caption{Undriven twisted MoTe$_2$ homobilayer at $\theta=4^{\circ}$ and $V_{z}=-2$\,meV. (a) Out-of-plane layer pseudo-spin $S_{z}(\mathbf{r})$ over several moir\'{e} periods; the hexagon and rhombus mark the superlattice and one primitive cell. (b) Static valence bands along high-symmetry lines. (c) Point-by-point comparison at $\hbar\omega=26$\,meV of the Kubo and Floquet responses for $\mu=x,y,z$; the line is the identity.}
\label{fig:bands}
\end{figure}

We solve the moir\'{e} Hamiltonian and the driven problem in the Floquet picture\cite{zhang2026quantum,floquet-fu}, diagonalizing the quasienergy operator on the mBZ and time-averaging the layer pseudo-spin (Supplemental
Material, Sec.~\ref{sec:SM:model} ).  Figure~\ref{fig:bands} shows the undriven system at $V_{z}=-2$\,meV: the local out-of-plane layer pseudo-spin $S_{z}(\mathbf{r})$ inside the moir\'{e} cell, and the static moir\'{e} valence bands along the high-symmetry lines of the mBZ.  The topmost band is separated from the remaining valence bands over the whole mBZ and its states are almost fully layer polarized: $S_{z}(\mathbf{r})$ changes sign across the lines on which the out-of-plane pseudo-field $\Delta_{z}(\mathbf{r})$ vanishes, and it is this pattern that the in-plane drive is going to modulate.

The Kubo response of Eq.~(\ref{eq:kubo}) is compared point by point with the first harmonic of the direct Floquet solution in Fig.~\ref{fig:bands}(c). The two describe the same object once the driven response is taken in the linear-response limit, its Sambe first harmonic conjugated into the convention of Eq.~(\ref{eq:kubo}); they then coincide at every $\mathbf{k}$ and every $\omega$, identifying the geometric tensor with the response of the driven band.


The single-valley Hamiltonian without out-of-plane displacement field $V_z$ has the point group $D_{3}$, generated by the threefold rotation $C_{3z}$ and by the twofold rotation $C_{2y}$ that exchanges the two layers, while time reversal exchanges the valleys \cite{Wu2019}.  Because the response tensor is a property of the unperturbed Hamiltonian, it is invariant under every element of the group. For the channel $\mu=z$ this has an immediate consequence: $\tau_{z}$ is invariant under $C_{3z}$, whereas the in-plane field is a genuine two-dimensional vector, and the product of the trivial representation with the vector representation contains no invariant.  Hence $A^{zx}=A^{zy}=0$, the \emph{spatially averaged} layer polarization has no linear response to an in-plane field 
(Supplemental Material, Sec.~\ref{sec:SM:rules} ). However, the momentum-resolved response is a \emph{local} quantity, and is therefore unconstrained: the drive imprints a moir\'{e}-periodic texture that integrates to zero over the mBZ.

Thus, an in-plane drive writes an \emph{intracell layer-polarization pattern} into the flat band: the layer character is \emph{patterned} rather than uniformly shifted, and the drive produces a distribution within a single moir\'{e} unit cell. Because $\tau_{z}$ compares the weights of the two layers at fixed total density, the electron density of one layer changes by $+\Delta S_{z}(\mathbf{r})/2$ and that of the other by $-\Delta S_{z}(\mathbf{r})/2$, where $\Delta S_{z}(\mathbf{r})=\sum_{\mathbf{G}}e^{i\mathbf{G}\cdot\mathbf{r}}\,
\Delta\langle u_{\mathbf{k}}|\tau_{z}|u_{\mathbf{k}+\mathbf{G}}\rangle $ is the Fourier transform of the momentum-space texture onto the moir\'{e} cell; its cell average, the net layer polarization, is the quantity that $A^{zx}=A^{zy}=0$ constrains.  The corresponding local out-of-plane dipole density is $\delta P_{z}(\mathbf{r})=(ed/2)\Delta S_{z}(\mathbf{r})/A_{\mathrm{cell}}$ (Supplemental Material, Eq.~(\ref{eq:SM:polarization}) ).  A single layer therefore registers the drive directly, as a moir\'{e}-periodic modulation of its electron density, of the interlayer potential it experiences, and of its layer-sensitive optical selection rules: a scanning tunnelling measurement of one layer images that density, and therefore sees the pattern of the layer polarization, while the net polarization of the sample stays fixed.

A spatially averaged effect appears at higher order in the field.  At $V_{z}=0$ the Hamiltonian is invariant under $C_{2y}$, which acts on the layer polarization as $\tau_{z}\to-\tau_{z}$ and maps a linearly polarized drive along $\varphi$ onto the drive along $-\varphi$ up to a half-period shift.  The rectified (static) component of $P_{z}$ is quadratic in the field at leading order and isotropic by $C_{3z}$, so that the only available invariant is $\langle E_{x}^{2}+E_{y}^{2}\rangle$, which is even under $C_{2y}$ and must therefore vanish; this excludes the rectified response through fourth order in the field, and the leading nonvanishing term at $V_{z}=0$ appears at sixth order (Supplemental Material, Sec.~\ref{sec:SM:rules} ).  Turning on $V_{z}$ breaks the layer exchange and gives
\begin{equation}
P_{z}^{(\mathrm{rec})}=c_{2}\,V_{z}
\bigl\langle E_{x}^{2}+E_{y}^{2}\bigr\rangle
+\mathcal{O}(V_{z}^{3},E_{0}^{4}):
\label{eq:rectified}
\end{equation}
a static layer polarization induced by an in-plane field, whose sign follows the displacement field and whose magnitude grows quadratically with the drive. The same reasoning applies to a static in-plane field.

A circularly polarized drive is odd under $C_{2y}$, which maps it onto the opposite helicity rather than onto itself, but the combination of $C_{2y}$ with time reversal leaves the drive invariant and is a symmetry of the layer-symmetric problem: the averaged response therefore vanishes at $V_{z}=0$ for both drives and for every polarization (Supplemental Material, Sec.~\ref{sec:SM:rules}), and the helicity-odd channel appears with the same displacement field that switches on the rectified one.  Time reversal relates the two valleys and yields, for the helicity-odd part of the valley-resolved response,
\begin{equation}
\begin{split}
P_{z}^{(+K)}(h)-P_{z}^{(+K)}(-h)
&=-P_{z}^{(-K)}(h)+P_{z}^{(-K)}(-h)\\
&=2c_{3}\,h\,V_{z}\,E_{0}^{2}+\mathcal{O}(V_{z}^{3},E_{0}^{4}),
\end{split}
\label{eq:helicity}
\end{equation}
where $h$ is the helicity.  The two valleys acquire \emph{opposite} helicity-odd layer polarizations, so that a valley-resolved measurement sees a helicity-controlled layer polarization while the net dipole of the sample vanishes.

\begin{figure*}[bhtp]
\centering
\includegraphics[width=0.9\textwidth]{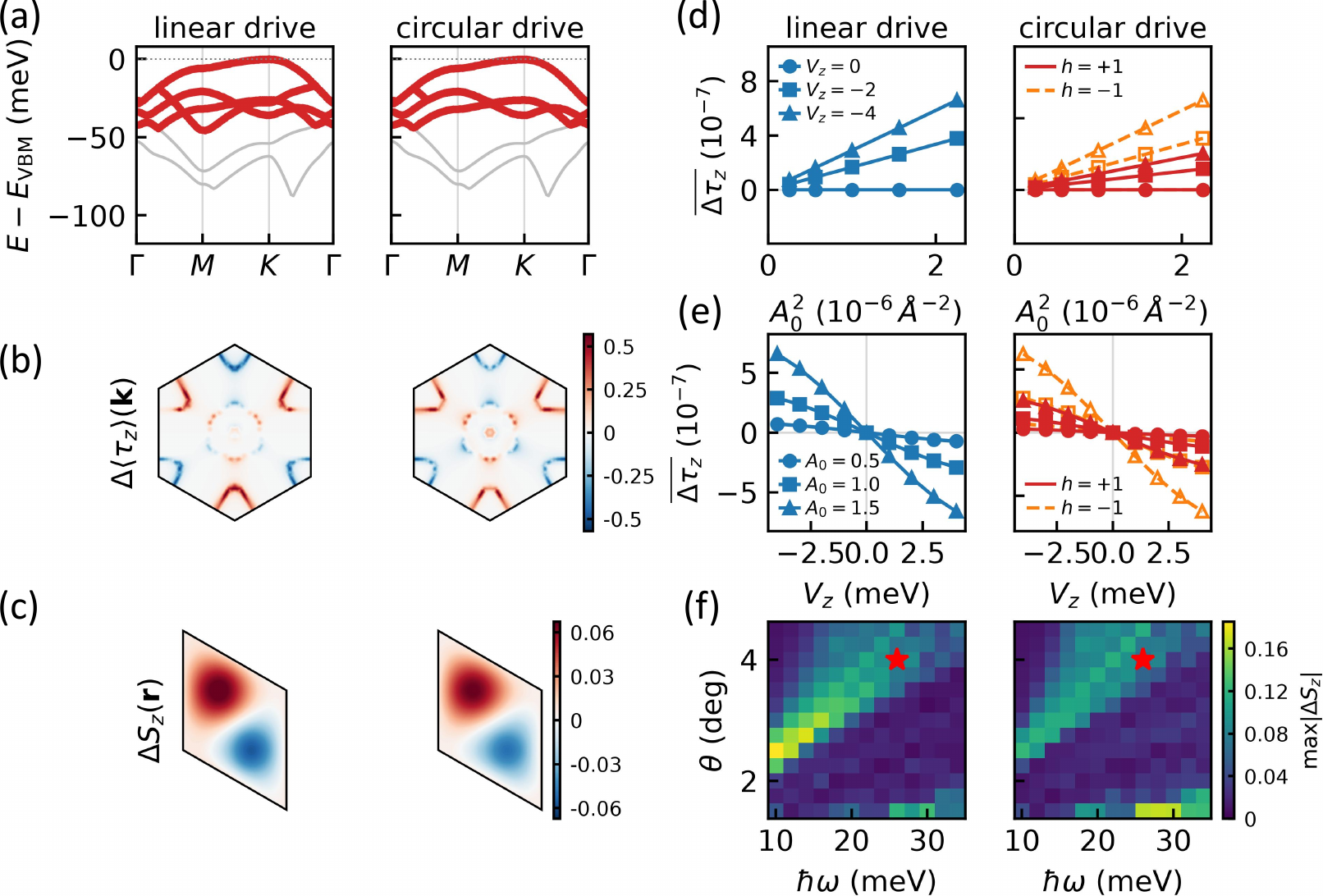}
\caption{Driven layer pseudo-spin at $\theta=4^{\circ}$. Left, at $\hbar\omega=26$\,meV and $A_{0}=0.005$\,\AA$^{-1}$ for linear (left column) and circular (right column) drive: (a) quasi-energy bands (symbols) and static bands (grey lines); (b) momentum-resolved $\Delta\langle\tau_{z}\rangle$ on the mBZ; (c) corresponding intracell pattern $\Delta S_{z}(\mathbf{r})$. Right, at the sub-gap $\hbar\omega=1.5$\,meV: (d) cell-averaged $\overline{\Delta\tau_{z}}$ versus $A_{0}^{2}$ for $V_{z}=0,-2,-4$\,meV; (e) $\overline{\Delta\tau_{z}}$ versus $V_{z}$ for $A_{0}=0.5,1.0,1.5\times10^{-3}$\,\AA$^{-1}$; solid and dashed lines in the circular-drive panels denote $h=\pm1$. (f) Maximum $|\Delta S_{z}(\mathbf{r})|$ over $(\theta,\hbar\omega)$ at $V_{z}=0$ and $A_{0}=0.005$\,\AA$^{-1}$; the star marks the parameters used above.}
\label{fig:overview}
\end{figure*}

For a linearly polarized drive the fundamental vanishes by $A^{zx}=A^{zy}=0$ and the even harmonics by the $C_{2y}$ argument, so that the leading averaged response is the third harmonic, that is, the component of $P_{z}$ that oscillates at three times the frequency of the drive,
$
P_{z}^{(3\omega)}(t)\propto E_{0}^{3}\cos3\varphi\,\cos(3\omega t)
+\mathcal{O}(E_{0}^{5}).
\label{eq:thirdharmonic}
$

These results hold for every $C_{3}$-symmetric moir\'{e} system, and it reflects the representation content of the layer degree of freedom.  A spin transforms as a vector under the threefold rotation, so its product with the in-plane field contains an invariant and a spin polarization follows at linear order \cite{ZeemanOrigin,ZeemanQG}; the layer polarization is a scalar, its product with the field is a vector, and this product carries no invariant. The momentum-resolved $z^{z\nu}_{nm}(\mathbf{k})$ is unconstrained, so its imaginary part, the Zeeman--Berry curvature of the layer pseudo-spin, is a genuine geometric property of the flat band that the in-plane drive exposes directly.

\emph{Calculated results about layer-pseudo-spin response.}  
Figure~\ref{fig:overview} collects the driven results. Panels (a)--(c) compare the linear and circular drives at the ridge energy $\hbar\omega=26$\,meV, where $|\Delta S_{z}|$ is maximal for $\theta=4^{\circ}$: quasi-energy bands, the momentum-resolved change $\Delta\langle\tau_{z}\rangle(\mathbf{k})$, and the intracell pattern $\Delta S_{z}(\mathbf{r})$. Panels (d) and (e) give the cell-averaged response at the sub-gap energy $\hbar\omega=1.5$\,meV, versus $A_{0}^{2}$ for three values of $V_{z}$ and versus $V_{z}$ for three drive amplitudes. Panel (f) maps the maximum real-space response over $(\theta,\hbar\omega)$; the response follows the ridge $\hbar\omega\simeq6.6\,\mathrm{meV/deg}\times\theta$ and the star marks the parameters used above.

For the linear drive $\Delta\langle\tau_{z}\rangle$ is concentrated near $\Gamma$ and reverses sign within the mBZ: its peak is a large fraction of the pseudo-spin, while its zone average is two orders of magnitude smaller.  For the circular drive the response is an extended ridge along a symmetry line of the mBZ, the single-valley image of the helicity-odd channel of Eq.~(\ref{eq:helicity}), and the other valley contributes the opposite value.  Fourier transforming onto the moir\'{e} cell makes the intracell pattern explicit in panel (c): the local change $\Delta S_{z}(\mathbf{r})$ is a moir\'{e}-periodic pattern of alternating sign whose intra-cell modulation is a few per cent of the saturation value of the pseudo-spin.  Thus, a scanning tunnelling measurement of one layer images it directly.

The cell average, panels (d) and (e), is the quantity that $A^{zx}=A^{zy}=0$ constrains at linear order.  It vanishes identically at $V_{z}=0$ for both drives, as required by the layer-exchange argument of Sec.~\ref{sec:SM:rules}, and it becomes finite once $V_{z}$ breaks that symmetry: this is the rectified channel of Eq.~(\ref{eq:rectified}), even in the field and odd in $V_{z}$.  The circular drive carries in addition the helicity-odd part of Eq.~(\ref{eq:helicity}), also odd in $V_{z}$, which splits the two helicities in the circular-drive panels (d) and (e).  The average is taken on a uniform $\mathbf{k}$ grid on the mBZ torus, for which the trapezoid rule for the periodic integrand is spectrally accurate: the \emph{undriven} cell average then converges to five digits and is linear in $V_{z}$ to better than a per cent over the range of the panels.  The \emph{driven} average is a residual of a cancellation of momentum-resolved responses of order one, and it converges only while the driven flat band stays isolated from all of its photon replicas: the moir\'{e} gap between the two topmost valence bands is smallest at the centre of the mBZ, $\simeq5$\,meV at $\theta=4^{\circ}$, and the drive energy must be low enough that neither it nor its first multiples reach that minimum.  Panels (d) and (e) are therefore evaluated at the sub-gap drive energy $\hbar\omega=1.5$\,meV, for which the first resonance of the comb $n\hbar\omega=\Delta(\mathbf{k})$ is at $n=4$, and at $A_{0}\le1.5\times10^{-3}$\,\AA$^{-1}$.  Panel (d) then establishes the $A_{0}^{2}$ growth and the vanishing at $V_{z}=0$, while panel (e) shows the oddness in $V_{z}$ of both the rectified and the helicity-odd parts of the average.

Scanning the drive parameters, as shown in panel (f), the response grows with the twist angle and is largest in a broad window where the moir\'{e} bands are flattest.  The intra-cell amplitude is systematically smaller than the momentum-space one, since a response that oscillates in sign in momentum space has a reduced Fourier transform. As shown in Fig.3, the maximum $\Delta S_{z}(\mathbf{r})=S_{z}(\mathbf{r})_{\mathrm{driven}}-S_{z}(\mathbf{r})_{\mathrm{static}}$ can reach about 20\% under linear or circular drive, which can be detected by scanning tunneling microscopic experiments. 

\begin{figure}[htbp]
\centering
\includegraphics[width=\columnwidth]{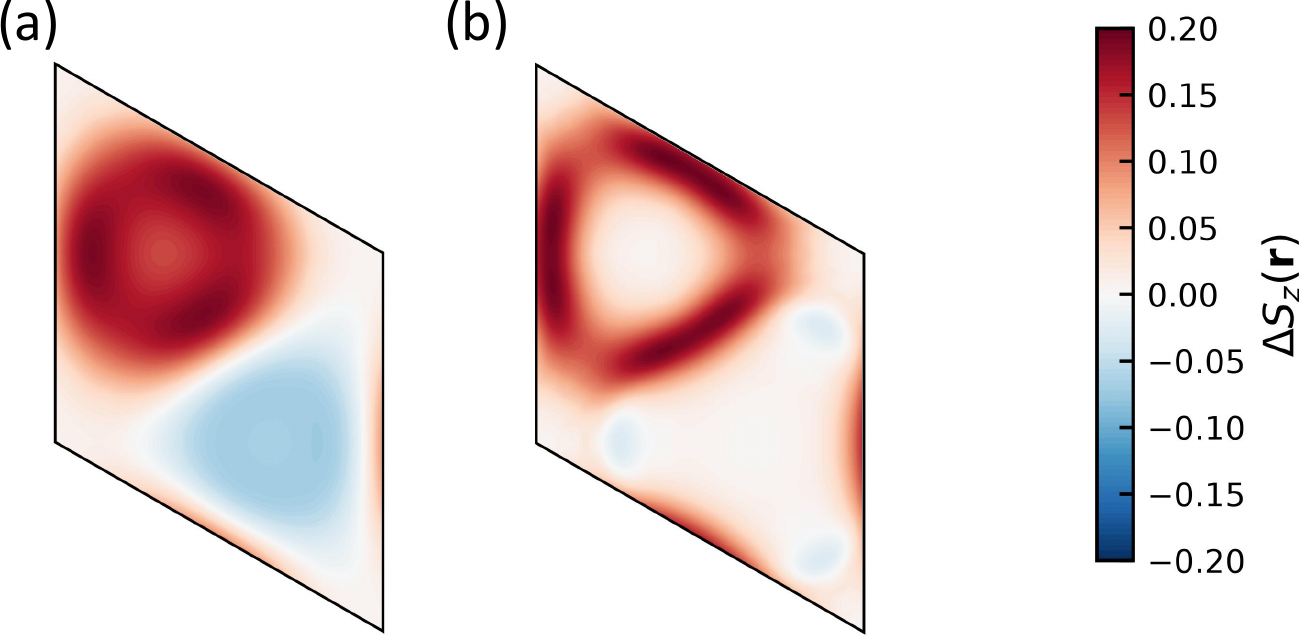}
\caption{Real-space layer-pseudo-spin response at the strongest points of the $(\theta,\hbar\omega)$ scan. (a) and (b) maps show $\Delta S_{z}(\mathbf{r})$ in one moir\'{e} cell for the linear- and circular-drive maxima, respectively. Within the scanned interval
$1.5^{\circ}\le\theta\le4.5^{\circ}$ and
$10\,\mathrm{meV}\le\hbar\omega\le34\,\mathrm{meV}$, the linear-drive maximum occurs at $(\theta,\hbar\omega)=(2.50^{\circ},10\,\mathrm{meV})$ with $\max|\Delta S_{z}|=0.189$, while the circular-drive maximum occurs at $(1.50^{\circ},28\,\mathrm{meV})$ with $\max|\Delta S_{z}|=0.197$.}
\label{fig:SM:scanmaxima}
\end{figure}


\emph{More discussions.}  There are three features of the intracell pattern are directly measurable. First, its period is the moir\'{e} period: the modulation of the layer character has the scale of the superlattice cell, and it should appear in the layer-resolved local density of states of scanning tunnelling spectroscopy and in the moir\'{e}-periodic modulation of the layer-sensitive optical selection rules. Second, the effect uses a purely in-plane field, which can be switched on without a displacement field and without touching the out-of-plane structure of the band.  Third, the nonlinear channels carry their own fingerprints: the rectified response of Eq.~(\ref{eq:rectified}) is linear in $V_{z}$ and even in the field, the response to a circular drive, Eq.~(\ref{eq:helicity}), is odd in the helicity and opposite in the two valleys, and the third harmonic of Eq.~(\ref{eq:thirdharmonic}) is identified by its frequency and by the sixfold pattern of its amplitude.


The picture that emerges is that of an in-plane field as a tool that \emph{shapes} rather than \emph{shifts} the layer character of a moir\'{e} flat band: through the quantum geometry of the layer pseudo-spin, the layer degree of freedom can be modulated on the moir\'{e} scale while the out-of-plane structure of the flat bands, on which their correlated states are built, is left intact.  Given that twisted TMD homobilayers are the platform of choice for moir\'{e} ferroelectricity and for the correlated states of moir\'{e} flat bands, and that the effect exists in every $C_{3}$-symmetric moir\'{e} system, these symmetry constraints should be a useful guide for experiments that manipulate the layer degree of freedom with in-plane fields.

\emph{Acknowledgments.} This work is supported by the National Key Research and Development Program of China (Grant No. 2024YFA1410300), the National Natural Science Foundation of China (Grant No. 12304217), the Natural Science Foundation of HunanProvince (Grant No. 2025JJ60002), the Fundamental Research Funds for the Central Universities from China (Grant No. 531119200247).

\onecolumngrid
\clearpage
\renewcommand{\thesection}{S\arabic{section}}
\renewcommand{\theequation}{S\arabic{equation}}
\renewcommand{\thefigure}{S\arabic{figure}}
\renewcommand{\thetable}{S\arabic{table}}
\setcounter{section}{0}
\setcounter{equation}{0}
\setcounter{figure}{0}
\setcounter{table}{0}

\begin{center}
{\large\bfseries Supplementary Material for}\\[6pt]
{\large\bfseries Zeeman Quantum Geometry of the Layer Pseudo-spin}
\end{center}

\newcounter{smsec}
\renewcommand{\thesmsec}{S\arabic{smsec}}
\newcommand{\smsection}[2]{%
\refstepcounter{smsec}\section*{\thesmsec\quad #1}\label{#2}}

\smsection{Continuum model and numerical method}{sec:SM:model}

Following Ref.~\cite{Wu2019} we describe the valence states of a twisted MoTe$_2$ homobilayer in the $+K$ valley by a two-band Hamiltonian in the layer pseudo-spin basis; the states of the $-K$ valley follow from time reversal.  We keep one band per valley, the topmost valence band: Ising spin--orbit coupling locks the spin to the valley, so that the spin is frozen, the continuum model is spinless and no spin--orbit term appears in it explicitly.  Writing $\tau_{z}=+1$ ($-1$) for the bottom (top) layer, the layer-dependent potentials and the interlayer tunnelling amplitude are
\begin{align}
\Delta_{\mathfrak{b},\mathfrak{t}}(\mathbf{r})
&=2V\sum_{j=1,3,5}\cos\!\left(\mathbf{b}_{j}\!\cdot\!\mathbf{r}
\pm\psi\right),
\label{eq:SM:potentials}\\
\Delta_{T}(\mathbf{r})
&=w\left(1+e^{-i\mathbf{G}_{2}\cdot\mathbf{d}}
+e^{-i\mathbf{G}_{3}\cdot\mathbf{d}}\right),
\label{eq:SM:tunnelling}
\end{align}
where the upper (lower) sign in Eq.~(\ref{eq:SM:potentials}) refers to the bottom (top) layer and $\mathbf{d}(\mathbf{r})=\theta\,\hat{z}\times\mathbf{r}$ is the local interlayer displacement.  The reciprocal vectors and the moir\'{e} length scales are
\begin{equation}
\mathbf{b}_{j}=\theta\,\hat{z}\times\mathbf{G}_{j},\qquad
\mathbf{G}_{j}=\hat{R}\!\left(\frac{(j-1)\pi}{3}\right)\mathbf{G}_{1},\qquad
\mathbf{G}_{1}=\frac{4\pi}{\sqrt{3}a_{0}}\,\hat{y},
\label{eq:SM:Gvectors}
\end{equation}
\begin{equation}
a_{M}=\frac{a_{0}}{\theta},\qquad
|\mathbf{b}_{j}|=\frac{4\pi}{\sqrt{3}\,a_{M}},\qquad
|\boldsymbol{\kappa}_{\pm}|=\frac{4\pi}{3\,a_{M}},
\label{eq:SM:moirelength}
\end{equation}
with $\hat{R}(\phi)$ a counterclockwise rotation by $\phi$; $\boldsymbol{\kappa}_{\pm}$ are the mBZ corners at which the Dirac points of the two layers are folded.  We use the monolayer lattice constant $a_{0}=3.52$\,\AA{} of MoTe$_2$, so that $a_{M}\simeq5.0$\,nm at $\theta=4^{\circ}$.  Note that $a_{0}$ enters only through Eq.~(\ref{eq:SM:Gvectors}) and the moir\'{e} period in Eq.~(\ref{eq:SM:moirelength}); using the monolayer rather than the moir\'{e} lattice constant in the Dirac-point positions of Eq.~(\ref{eq:SM:moirelength}) would overestimate $|\boldsymbol{\kappa}_{\pm}|$ by a factor of $\sqrt{3}/\theta$.  Because $\psi=-89.6^{\circ}\simeq-90^{\circ}$, the layer-symmetric component $\Delta_{+}\propto\cos\psi$ is strongly suppressed while $\Delta_{z}\propto\sin\psi$ is large: the layer pseudo-spin precesses in a predominantly out-of-plane pseudo-magnetic field, which is the origin of the pronounced layer-pseudo-spin texture discussed below.  Two remarks on the coupling to a uniform in-plane field are in order.  First, for a uniform field the minimal coupling of Eq.~(\ref{eq:drive}) of the main text acts on the intralayer kinetic term; because the moir\'{e} potential and the interlayer tunnelling are functions of the local displacement $\mathbf{d}(\mathbf{r})$, the field also enters the tunnelling amplitude through a Peierls phase $\exp[i\mathbf{A}\cdot\mathbf{d}(\mathbf{r})]$.  For the amplitudes used here the maximum phase acquired by the tunnelling term is $A_{0}|\mathbf{d}|\simeq A_{0}\theta a_{M}/\sqrt{3}\simeq10^{-2}$; since this phase and the intrinsic moir\'{e} phases $\mathbf{G}_{j}\cdot\mathbf{d}$ are both proportional to $|\mathbf{d}|$, their ratio is simply $A_{0}/|\mathbf{G}_{j}|\simeq0.03$.  This off-diagonal coupling (which acts in the $(\tau_{x},\tau_{y})$ channel, i.e.\ in the same channel as $\Delta_{T}$) is not included in the present calculation and deserves a separate study.  Since it preserves the point group, it does not affect the symmetry analysis of Sec.~\ref{sec:SM:rules} and cannot contribute to the net layer polarization.  Second, we neglect the out-of-plane component of the field, so that the displacement potential $V_{z}$ is not modulated by the drive.

The moir\'{e} Hamiltonian of Eq.~(\ref{eq:model}) of the main text is solved by plane-wave expansion in the moir\'{e} reciprocal lattice.  Under the drive the time-dependent Schr\"{o}dinger equation is solved in the Floquet (extended Hilbert space) picture, $(H(t)-i\partial_{t})|u\rangle=\varepsilon|u\rangle$, by expanding the periodic part in Floquet sidebands $e^{-im\omega t}$ with $m=-M,\dots,M$, and diagonalising the resulting quasienergy operator for each $\mathbf{k}$ of the mBZ.  Convergence is checked with respect to the number of moir\'{e} harmonics, the number of sidebands and the $\mathbf{k}$ mesh.  The driven states are characterised by the time-averaged layer pseudo-spin
\begin{equation}
\label{eq:SM:timeaverage}
\langle\boldsymbol{\tau}\rangle(\mathbf{k})
=\frac{1}{T}\int_{0}^{T}\!dt\,
\langle\psi_{\mathbf{k}}(t)|\boldsymbol{\tau}|\psi_{\mathbf{k}}(t)\rangle
=\sum_{m}\langle u^{(m)}_{\mathbf{k}}|\boldsymbol{\tau}|u^{(m)}_{\mathbf{k}}\rangle,
\end{equation}
and we quote the change $\Delta\langle\tau_{z}\rangle$ with respect to the undriven state, resolved either in momentum space or, after Fourier transformation on the mBZ to the moir\'{e} cell, in real space,
\begin{equation}
\label{eq:SM:realspace}
\Delta S_{z}(\mathbf{r})=\sum_{\mathbf{G}}
e^{i\mathbf{G}\cdot\mathbf{r}}\,
\Delta\langle u_{\mathbf{k}}|\tau_{z}|u_{\mathbf{k}+\mathbf{G}}\rangle,
\end{equation}
where $\mathbf{G}$ runs over the moir\'{e} reciprocal lattice and the matrix elements are taken in the plane-wave basis.

The quantity of interest in this work is the change of the layer polarization that a \emph{single layer} registers.  Because $\tau_{z}$ compares the weights of the two layers, the electron density of the layer with $\tau_{z}=+1$ changes by $+\Delta S_{z}(\mathbf{r})/2$ and that of the other layer by $-\Delta S_{z}(\mathbf{r})/2$; $\Delta S_{z}(\mathbf{r})$ is therefore directly the local, layer-resolved signal of the drive, and it is the natural observable of layer-sensitive probes such as layer-contrast ARPES or layer-resolved scanning tunnelling spectroscopy, as well as the source of a locally modulated interlayer potential difference.  The accompanying local out-of-plane dipole density is
\begin{equation}
\label{eq:SM:polarization}
\delta P_{z}(\mathbf{r})=\frac{ed}{2}\,
\frac{\Delta S_{z}(\mathbf{r})}{A_{\mathrm{cell}}},\qquad
A_{\mathrm{cell}}=a_{M}^{2},
\end{equation}
so that the intra-cell modulation of Fig.~\ref{fig:overview}(e) of the main text corresponds to a local modulation of the dipole density of $5\times10^{-14}$\,C\,m$^{-2}$ ($5\times10^{-12}\,\mu$C\,cm$^{-2}$) for $a_{M}=5.0$\,nm and $d=7$\,\AA.

The $\mathbf{G}=0$ Fourier component of Eq.~(\ref{eq:SM:realspace}) is the cell average of $\Delta S_{z}$, i.e.\ the change of the \emph{net} layer polarization.  It is not the quantity we focus on.  In the layer-symmetric system it is required to vanish identically by the symmetry arguments of Sec.~\ref{sec:SM:rules}; at $V_{z}\neq0$ it is the rectified response of Eq.~(\ref{eq:rectified}), and we evaluate it on a uniform grid on the mBZ torus, for which the trapezoid rule for the periodic integrand is spectrally accurate (the undriven average converges to five digits and is linear in $V_{z}$ to better than a per cent).  The driven average is nevertheless only indicative at $\hbar\omega=26$\,meV, because the drive energy then exceeds the moir\'{e} gap over most of the mBZ and the driven flat band is not isolated from its photon replicas, so that its momentum-resolved modulation varies sharply along the contour on which the two meet.  At the lower drive energy of panels (d) and (e) of Fig.~\ref{fig:overview} of the main text the driven bands are isolated and the average is converged (Sec.~\ref{sec:SM:numerics} ).  For reference, a cell average $\overline{\langle\tau_{z}\rangle}=10^{-2}$ would correspond to an interlayer dipole moment of $0.035\,e$\,\AA{} per moir\'{e} cell, i.e.\ an areal polarization of $2\times10^{-14}$\,C\,m$^{-2}$, which is small compared with the static polarizations induced by a displacement field; the single-layer modulations that we do report are finite point by point and are therefore the experimentally relevant quantity.

\smsection{Geometric construction of the Zeeman quantum geometry}{sec:SM:geometry}

We collect here the geometric construction that underlies Eqs.~(\ref{eq:ztensor})--(\ref{eq:kubo}) of the main text, following the construction of the Zeeman quantum geometry of Ref.~\cite{ZeemanOrigin}.  Let $|u_{n\mathbf{k}}\rangle$ be the cell-periodic valence state of band $n$, and let
\begin{equation}
U_{d\mathbf{k}}=e^{-i\,dk_\nu\hat{r}_\nu},\qquad
U_{d\boldsymbol{\theta}}=e^{-i\,d\theta_\mu\tau_\mu/2},
\label{eq:SM:Uoperators}
\end{equation}
where $\hat{r}_\nu=i\partial_{k_\nu}$ generates an infinitesimal momentum translation and $\tau_\mu/2$ generates an infinitesimal rotation in layer space.  Following Ref.~\cite{ZeemanOrigin} we generalise the quantum distance between two infinitesimally separated states to
\begin{equation}
ds^{2}=\bigl\|U_{d\boldsymbol{\theta}}U_{d\mathbf{k}}
|u_{n\mathbf{k}}\rangle-|u_{n\mathbf{k}}\rangle\bigr\|^{2},
\label{eq:SM:distance}
\end{equation}
which reduces to the conventional quantum distance when $d\boldsymbol{\theta}=0$.  Using $U_{d\boldsymbol{\theta}}U_{d\mathbf{k}}=\mathbf{1}+dk_\nu\partial_{k_\nu} -\tfrac{i}{2}d\theta_\mu\tau_\mu+\cdots$, the state whose norm is taken in Eq.~(\ref{eq:SM:distance}) differs from $|u_{n\mathbf{k}}\rangle$ by $dk_\nu|\partial_{k_\nu}u_{n\mathbf{k}}\rangle -\tfrac{i}{2}d\theta_\mu\tau_\mu|u_{n\mathbf{k}}\rangle$, and subtracting the band-diagonal pieces, which is what removes the dependence on the phase convention of $|u_{n\mathbf{k}}\rangle$ and on the layer-dependent phase of the two-component basis, leaves the gauge-invariant second-order form
\begin{equation}
ds^{2}=\sum_{m\neq n}\Bigl[\,
g^{\nu\nu'}_{nm}\,dk_\nu dk_{\nu'}
+\tfrac{1}{4}\Sigma^{\mu\mu'}_{nm}\,d\theta_\mu d\theta_{\mu'}
+2\,\mathcal{Q}^{\mu\nu}_{nm}\,d\theta_\mu dk_\nu\Bigr],
\label{eq:SM:distanceexpand}
\end{equation}
with
\begin{equation}
g^{\nu\nu'}_{nm}=a^{\nu}_{nm}a^{\nu'}_{mn},\qquad
\Sigma^{\mu\mu'}_{nm}=s^{\mu}_{nm}s^{\mu'}_{mn},\qquad
\mathcal{Q}^{\mu\nu}_{nm}=\operatorname{Re}z^{\mu\nu}_{nm}.
\label{eq:SM:gQSigma}
\end{equation}
The first term is the ordinary quantum metric of the Bloch states, the second is its layer analogue, and the third is \emph{mixed} in the two: it contains one momentum derivative and one layer derivative, and it is generated entirely by the tensor $z^{\mu\nu}_{nm}=a^{\nu}_{nm}s^{\mu}_{mn}$ of Eq.~(\ref{eq:ztensor}) of the main text.  Its real part is the Zeeman quantum metric that appears in Eq.~(\ref{eq:SM:distanceexpand}); its imaginary part,
\begin{equation}
\mathcal{Z}^{\mu\nu}_{nm}=2\,\operatorname{Im}z^{\mu\nu}_{nm}
=i\bigl(a^{\nu}_{mn}s^{\mu}_{nm}-a^{\nu}_{nm}s^{\mu}_{mn}\bigr),
\label{eq:SM:Zdef}
\end{equation}
does not enter the quantum distance, which is symmetric under the interchange of the two infinitesimal displacements, but it controls the antisymmetric (quadrature) part of the response derived below.  Equations~(\ref{eq:SM:Uoperators})--(\ref{eq:SM:Zdef}) are the layer-pseudo-spin analogues of Eqs.~(3)--(6) of Ref.~\cite{ZeemanOrigin}, and they reproduce the definitions of Eq.~(\ref{eq:ztensor}) of the main text.  One difference deserves emphasis: for the spin, $U_{d\boldsymbol{\theta}}$ rotates a physical external degree of freedom, whereas for the layer pseudo-spin it is a formal device that generates the mixed term; the tensor it produces, $z^{\mu\nu}_{nm}$, is nevertheless the object that governs the response, as we now show.

\smsection{Minimal two-band model}{sec:SM:minimal}

We collect the algebra behind Eqs.~(\ref{eq:minimal})--(\ref{eq:minimalQZ}) of the main text.  Write the two-band Hamiltonian as
\begin{equation}
H_{\min}=\varepsilon(\mathbf{k})\,\tau_{0}
+\boldsymbol{\Delta}(\mathbf{k})\cdot\boldsymbol{\tau},
\qquad
\boldsymbol{\Delta}=\bigl(\Delta_{\parallel}\cos\gamma,\;
\Delta_{\parallel}\sin\gamma,\;\Delta_{z}\bigr),
\label{eq:SM:minimalH}
\end{equation}
with $E=|\boldsymbol{\Delta}|$, so that $\cos\theta=\Delta_{z}/E$ and $\sin\theta=\Delta_{\parallel}/E$.  Its eigenstates are
\begin{equation}
|u_{+}\rangle=\binom{\cos(\theta/2)}{\sin(\theta/2)e^{i\gamma}},
\qquad
|u_{-}\rangle=\binom{-\sin(\theta/2)e^{-i\gamma}}{\cos(\theta/2)},
\label{eq:SM:minimalstates}
\end{equation}
which give $\langle\tau_{z}\rangle_{\pm}=\pm\cos\theta$ and $\langle\tau_{x,y}\rangle_{\pm}=\pm\sin\theta(\cos\gamma,\sin\gamma)$, the texture quoted in the main text.  The interband Berry connection and the layer matrix element are
\begin{equation}
a^{\nu}_{+-}=-\tfrac{1}{2}e^{-i\gamma}
\bigl(\sin\theta\,\partial_{k_{\nu}}\gamma+i\,\partial_{k_{\nu}}\theta\bigr),
\qquad
s^{z}_{-+}=-\sin\theta\,e^{i\gamma},
\label{eq:SM:minimalas}
\end{equation}
so that the mixed tensor $z^{z\nu}_{+-}=a^{\nu}_{+-}s^{z}_{-+}$ is
\begin{equation}
z^{z\nu}_{+-}=\tfrac{1}{2}\sin^{2}\!\theta\,\partial_{k_{\nu}}\gamma
+\tfrac{i}{2}\sin\theta\,\partial_{k_{\nu}}\theta
=\mathcal{Q}^{z\nu}+\tfrac{i}{2}\mathcal{Z}^{z\nu},
\label{eq:SM:minimalz}
\end{equation}
which is Eqs.~(\ref{eq:minimalz}) and (\ref{eq:minimalQZ}) of the main text; the second form of Eq.~(\ref{eq:minimalQZ}) of the main text follows from $\partial_{k_{\nu}}\cos\theta=-\sin\theta\,\partial_{k_{\nu}}\theta$.  For the third-harmonic pseudo-field of Eq.~(\ref{eq:minimalfield}) of the main text, $\gamma=3\varphi+\gamma_{0}$, so that $\partial_{k_{\nu}}\gamma=3(-\sin\varphi,\cos\varphi)_{\nu}/k$ and $\partial_{k_{\nu}}\theta\propto k_{\nu}$.  Both $\mathcal{Q}^{z\nu}$ and $\mathcal{Z}^{z\nu}$ are therefore the components of an in-plane vector multiplied by a threefold-invariant scalar, and their integrals over a threefold-symmetric Brillouin zone vanish identically; this is the minimal-model version of Rule~1 of Sec.~\ref{sec:SM:rules}.

\smsection{Derivation of the response coefficients}{sec:SM:kubo}

We take the uniform in-plane field of Eq.~(\ref{eq:drive}) of the main text, $\mathbf{E}(t)=\mathbf{E}\cos\omega t$, and write its coupling to the electronic states in the length gauge through the interband position operator,
\begin{equation}
H'(t)=-E_\nu(t)\,\hat{r}_\nu,\qquad
\langle u_{m\mathbf{k}}|\hat{r}_\nu|u_{n\mathbf{k}}\rangle=a^{\nu}_{mn}
\quad(m\neq n),
\label{eq:SM:perturbation}
\end{equation}
whose Fourier components at the two frequencies $\omega_s=\pm\omega$ of the real drive are $H'(\omega_s)=-\tfrac12E_\nu\hat{r}_\nu$.  The density matrix $\rho=\rho_0+\delta\rho$ obeys the quantum Liouville equation $i\partial_t\delta\rho=[H_0,\delta\rho]+[H',\rho_0]$, and with $(\rho_0)_{mn}=f_n\delta_{mn}$ together with the ansatz $\delta\rho_{mn}(t)=\delta\rho_{mn}(\omega_s)e^{-i\omega_st}$ it gives
\begin{equation}
\delta\rho_{mn}(\omega_s)=
\frac{(f_n-f_m)H'_{mn}(\omega_s)}{\omega_s-\varepsilon_{mn}+i\eta},
\qquad
\varepsilon_{mn}=\varepsilon_m-\varepsilon_n,
\label{eq:SM:drho}
\end{equation}
with $i\eta$ the adiabatic (retarded) factor.  The layer pseudo-spin is $S^{\mu}(t)=\sum_{mn}\rho_{mn}(t)s^{\mu}_{nm}$, so that to first order
\begin{equation}
\delta S^{\mu}(t)=\sum_{\omega_s=\pm\omega}\sum_{mn}
\delta\rho_{mn}(\omega_s)s^{\mu}_{nm}e^{-i\omega_st}.
\label{eq:SM:dS}
\end{equation}
Inserting Eq.~(\ref{eq:SM:drho}) with $H'_{mn}(\omega_s)=-\tfrac12E_\nu a^{\nu}_{mn}$, and using $a^{\nu}_{mn}s^{\mu}_{nm}=z^{*\mu\nu}_{nm}$ together with $f_{nm}=f_n-f_m$, the coefficient of $e^{-i\omega t}$ in Eq.~(\ref{eq:SM:dS}) gives
\begin{equation}
\delta S^{\mu}(t)=-\frac{E_\nu}{2}\sum_{mn}
\frac{f_{nm}z^{*\mu\nu}_{nm}}{\omega-\varepsilon_{mn}+i\eta}\,
e^{-i\omega t}+\mathrm{c.c.}
\label{eq:SM:dSfinal}
\end{equation}
With the definition $S^{\mu}(t)=\operatorname{Re}\,[\chi^{\mu\nu}E^\nu e^{-i\omega t}]$, the response function is therefore
\begin{equation}
\chi^{\mu\nu}(\omega)=-\int_{\mathrm{mBZ}}\!d^{2}k
\sum_{mn}\frac{f_{nm}z^{*\mu\nu}_{nm}}{\omega-\varepsilon_{mn}+i\eta},
\label{eq:SM:master}
\end{equation}
in which the sum runs over all ordered pairs of bands.  We finally group the two members of each band pair, using $z^{*\mu\nu}_{nm}=z^{\mu\nu}_{mn}$ and $f_{mn}=-f_{nm}$: writing $\varepsilon_{mn}=\varepsilon_n-\varepsilon_m>0$ for the pair in which $n$ is the upper and $m$ the lower band, the resonant and antiresonant members carry the denominators $\omega-\varepsilon_{mn}+i\eta$ and $\omega+\varepsilon_{mn}+i\eta$, respectively, so that Eq.~(\ref{eq:SM:master}) becomes
\begin{equation}
\chi^{\mu\nu}(\omega)=\int_{\mathrm{mBZ}}\!d^{2}k\sum_{n>m}f_{nm}\left[
\frac{z^{\mu\nu}_{nm}}{\omega-\varepsilon_{mn}+i\eta}
-\frac{z_{nm}^{*\,\mu\nu}}{\omega+\varepsilon_{mn}+i\eta}\right],
\label{eq:SM:kubo2}
\end{equation}
which is Eq.~(\ref{eq:kubo}) of the main text.  Equation~(\ref{eq:SM:kubo2}) is the layer-pseudo-spin counterpart of Eqs.~(7)--(8) of Ref.~\cite{ZeemanOrigin}: the interband dipole matrix element $a^{\nu}_{mn}$ plays the role of their interband velocity and the layer-pseudo-spin matrix element $s^{\mu}_{nm}$ the role of their spin matrix element, so that both response functions are built from the same mixed tensor $z^{\mu\nu}_{nm}$ of Eq.~(\ref{eq:ztensor}) of the main text.

Two remarks on Eq.~(\ref{eq:SM:master}) are in order.  First, we retain only the interband matrix elements of $\hat{r}_\nu$ in Eq.~(\ref{eq:SM:perturbation}), which are the gauge-invariant ones; the band-diagonal (intraband) part of the coupling and the diamagnetic (contact) contributions are omitted from the analytic formula, and they are included in the numerical Floquet solution of Sec.~\ref{sec:SM:model} used for the results of the main text.  Second, no assumption is made about the magnitude of the response, only that it is first order in the field.

\noindent\emph{Low-frequency expansion.}  For $\omega\ll\varepsilon_{mn}$ the two terms of Eq.~(\ref{eq:SM:kubo2}) can be expanded to first order in $\omega/\varepsilon_{mn}$,
\begin{equation}
\begin{split}
\frac{z}{\omega-\varepsilon+i\eta}
&\simeq-\frac{z}{\varepsilon}\left(1+\frac{\omega}{\varepsilon}\right)
=-\frac{z}{\varepsilon}-\frac{\omega z}{\varepsilon^{2}},\\
-\frac{z^{*}}{\omega+\varepsilon+i\eta}
&\simeq-\frac{z^{*}}{\varepsilon}
\left(1-\frac{\omega}{\varepsilon}\right)
=-\frac{z^{*}}{\varepsilon}+\frac{\omega z^{*}}{\varepsilon^{2}},
\end{split}
\label{eq:SM:expand}
\end{equation}
where $i\eta$ has been dropped since $\eta\ll\varepsilon$.  Adding the two terms and using $\mathcal{Q}^{\mu\nu}_{nm}=\operatorname{Re}z^{\mu\nu}_{nm}$ and $\mathcal{Z}^{\mu\nu}_{nm}=2\operatorname{Im}z^{\mu\nu}_{nm}$,
\begin{equation}
\chi^{\mu\nu}\simeq-\int_{\mathrm{mBZ}}\!d^{2}k\sum_{n>m}f_{nm}
\left[\frac{2\mathcal{Q}^{\mu\nu}_{nm}}{\varepsilon_{mn}}
+i\omega\frac{\mathcal{Z}^{\mu\nu}_{nm}}{\varepsilon_{mn}^{2}}\right]
=A^{\mu\nu}_{\cos}+iA^{\mu\nu}_{\sin},
\label{eq:SM:chi}
\end{equation}
which is Eqs.~(\ref{eq:Acos}) and (\ref{eq:Asin}) of the main text.  The real, $\omega$-independent part of the response is the quantum-metric contribution and the leading imaginary part is the Zeeman--Berry one, linear in $\omega$; this is the same decomposition as in Eqs.~(9)--(10) of Ref.~\cite{ZeemanOrigin}, with the Zeeman quantum metric and the Zeeman Berry curvature of Eqs.~(\ref{eq:SM:gQSigma}) and (\ref{eq:SM:Zdef}) in the place of the $N$ and $F$ coefficients defined there.  The expansion parameter is $\omega/\varepsilon_{mn}$, i.e.\ the ratio of the photon energy to the interband spacing, and the quadrature part is the leading term of the dissipative response of the driven system.

\smsection{Proofs of the selection rules}{sec:SM:rules}

The single-valley moir\'{e} Hamiltonian of Eq.~(\ref{eq:model}) of the main text has the point group $D_{3}$, generated by the threefold rotation $C_{3z}$ about $\hat{z}$ and the twofold rotation $C_{2y}$ about $\hat{y}$, the latter exchanging the two layers \cite{Wu2019}; time reversal $\hat{\mathcal{T}}$ exchanges the two valleys.  Since the linear-response tensor is a property of the unperturbed Hamiltonian, it must be invariant under every symmetry operation $g$ of the point group,
\begin{equation}
A^{\mu\nu}=D^{S}_{\mu\mu'}(g)^{-1}\,A^{\mu'\nu'}\,D^{E}_{\nu'\nu}(g),
\label{eq:SM:tensorrule}
\end{equation}
with $D^{E}=R$ the two-dimensional rotation acting on the in-plane components of the field and $D^{S}$ the representation acting on the pseudo-spin components.  The transformation properties are collected in Table~\ref{tab:SM:symmetry}.

\begin{table}[tbp]
\centering
\caption{Symmetry operations of the single-valley moir\'{e} model
[Eq.~(\ref{eq:model}) of the main text] and their action on the coordinates,
on the layer pseudo-spin and on the in-plane field.  The last column lists the
resulting selection rule.}
\label{tab:SM:symmetry}
\begin{ruledtabular}
\begin{tabular}{lllll}
Operation & $(\mathbf{r},\mathbf{k})$ & $(\tau_{x},\tau_{y},\tau_{z})$
& $\mathbf{E}_{\parallel}$ & Consequence \\
\hline
$C_{3z}$ & $(\mathcal{R}\mathbf{r},\mathcal{R}\mathbf{k})$
& $(\tau_{x},\tau_{y},\tau_{z})$ & $\mathcal{R}\mathbf{E}_{\parallel}$
& $A^{zx}=A^{zy}=0$ (Rule 1) \\
$C_{2y}$ & $(-x,y;-k_{x},k_{y})$ & $(\tau_{x},-\tau_{y},-\tau_{z})$
& $(-E_{x},E_{y})$ & rectified response $\propto V_{z}$ (Rule 2) \\
$\hat{\mathcal{T}}$ & $(\mathbf{r},-\mathbf{k})$
& $(\tau_{x},-\tau_{y},\tau_{z})$ & $\mathbf{E}_{\parallel}$
& $K\leftrightarrow K'$; helicity is odd \\
$C_{2y}\hat{\mathcal{T}}$ & $(-x,y;\bar{\mathbf{k}})$
& $(\tau_{x},\tau_{y},-\tau_{z})$ & $(-E_{x},E_{y})$
& $\overline{\Delta\tau_{z}}=0$ at $V_{z}=0$ (Rule 3) \\
$V_{z}$ & -- & $(0,0,\tau_{z})$ & -- & breaks $C_{2y}$ \\
\end{tabular}
\end{ruledtabular}
\end{table}

\noindent\emph{Rule 1.}  The layer pseudo-spin operators are invariant under $C_{3z}$, which acts on the spatial variables only; in the notation of Eq.~(\ref{eq:SM:tensorrule}) this means $D^{S}=\mathbf{1}$ for $C_{3z}$, whereas the in-plane field transforms in the two-dimensional representation $E$ with matrix $\mathcal{R}$ satisfying $\mathcal{R}^{3}=\mathbf{1}$.  Invariance of the response requires
\begin{equation}
A^{z\nu}=A^{z\nu'}\mathcal{R}_{\nu'\nu},
\end{equation}
i.e.\ the two-component object $(A^{zx},A^{zy})$ must be invariant under a rotation by $2\pi/3$.  The eigenvalues of a proper rotation of the plane are $e^{\pm2\pi i/3}$, neither of which is unity; the only real vector that is invariant is therefore the null vector, which gives $A^{zx}=A^{zy}=0$ of the main text.  Equivalently, $\tau_{z}$ carries the trivial representation $A_{1}$ of the point group $D_{3}$, the field carries $E$, and the product $A_{1}\otimes E=E$ contains no $A_{1}$.

\noindent\emph{Rule 2.}  (i) $C_{3z}$ invariance of the rank-two tensor $\Lambda_{\nu\nu'}$ defined by $P_{z}^{(\mathrm{dc})}=\Lambda_{\nu\nu'}\langle E^{\nu}E^{\nu'}\rangle$ requires $\Lambda=\mathcal{R}^{T}\Lambda\mathcal{R}$.  Writing $\Lambda=a\mathbf{1}+b\sigma_{x}+c\sigma_{z}$ and using $\mathcal{R}^{T}\sigma_{x}\mathcal{R}=-\tfrac12\sigma_{x} +\tfrac{\sqrt3}{2}\sigma_{z}$ and $\mathcal{R}^{T}\sigma_{z}\mathcal{R}=\tfrac{\sqrt3}{2}\sigma_{x} -\tfrac12\sigma_{z}$ gives $b=c=0$, so that $\Lambda=a\mathbf{1}$ and the quadratic response is isotropic in the plane.  (ii) At $V_{z}=0$ the Hamiltonian is invariant under $C_{2y}$, which acts on the field as $\mathbf{E}\to(-E_{x},E_{y})$ and on the layer polarization as $P_{z}\to-P_{z}$.  Invariance of the response then requires $a\langle E_{x}^{2}+E_{y}^{2}\rangle =-a\langle E_{x}^{2}+E_{y}^{2}\rangle$, hence $a=0$: with the layer-exchange symmetry intact the quadratic response vanishes, and with it the rectified response to a monochromatic linearly polarised drive.  (iii) The same counting applies at every even order in the field, where the response is the contraction of a $C_{3}$-invariant symmetric tensor $\Lambda^{(2k)}$ of rank $2k$ with $2k$ powers of the polarization vector.  Since $C_{3}$ acts on the polarization angle as $\varphi\to\varphi+2\pi/3$, only harmonics of $\varphi$ that are multiples of three can appear, and a rank-$2k$ tensor can generate only harmonics of degree at most $2k$.  For $2k\le4$ the response is therefore isotropic, i.e.\ a $\varphi$-independent multiple of $E_{0}^{2k}$, which is even under $C_{2y}$ and hence forbidden.  The first non-vanishing rectified term at $V_{z}=0$ is of sixth order, $P_{z}\propto E_{0}^{6}\sin6\varphi$, which is negligible at the field strengths considered here.  (iv) Breaking the layer-exchange symmetry by $V_{z}$ removes the parity argument of (ii): $V_{z}$ is odd under $C_{2y}$, so that $V_{z}\langle E_{x}^{2}+E_{y}^{2}\rangle$ is odd and allowed, while $C_{3}$ still forces the coefficient to be isotropic.  This yields Eq.~(\ref{eq:rectified}) of the main text.

\noindent\emph{Rule 3.}  A circularly polarised drive is odd under $C_{2y}$, which reverses both its handedness and the layer polarisation, $\tau_{z}\to-\tau_{z}$, and even under time reversal $\hat{\mathcal T}$, which exchanges the valleys.  The combination $C_{2y}\hat{\mathcal T}$ therefore leaves the drive invariant, and since both operations are symmetries of the layer-symmetric Hamiltonian at $V_{z}=0$, it is a symmetry of the \emph{driven} problem within a valley.  Its spatial part is the layer-exchanging mirror $\mathbf{k}\to\bar{\mathbf{k}}=(k_{x},-k_{y})$ and its pseudo-spin part is $\tau_{z}\to-\tau_{z}$, so that it maps the driven state at $\mathbf{k}$ onto the driven state at $\bar{\mathbf{k}}$ with the opposite layer character; the mBZ is invariant under the mirror, and the valley-resolved cell average therefore vanishes identically at $V_{z}=0$ for \emph{every} polarisation, circular ones included.  The helicity-odd channel is thus tied to the layer-exchange-breaking field $V_{z}$, exactly like the rectified channel of Rule 2.  The two operations separately give
\begin{equation}
C_{2y}:\;P_{z}^{(+K)}(h;V_{z})=-P_{z}^{(+K)}(-h;-V_{z}),\qquad
\hat{\mathcal T}:\;P_{z}^{(+K)}(h;V_{z})=P_{z}^{(-K)}(-h;V_{z}),
\label{eq:SM:valleyodd}
\end{equation}
whose combination gives, for the helicity-odd part, $P_{z}^{(-K)}(h)-P_{z}^{(-K)}(-h)=-[P_{z}^{(+K)}(h)-P_{z}^{(+K)}(-h)]$: the valley-odd structure of Eq.~(\ref{eq:helicity}) of the main text, with a vanishing sum over valleys, while the helicity-even part is valley-even and reduces to the rectified channel of Eq.~(\ref{eq:rectified}), in agreement with Rule 2.  The first relation also shows that the helicity-odd channel of a linearly polarised drive vanishes, that drive having $h=0$.

\noindent\emph{Rule 4 (third harmonic).} A monochromatic linearly polarised drive of frequency $\omega$ makes the response a periodic function of time with the period of the drive, so that it can be written as
\begin{equation}
P_{z}(t;\varphi)=\sum_{n=-\infty}^{\infty}P_{n}(\varphi)\,e^{-in\omega t},
\label{eq:SM:harmonics}
\end{equation}
where $\varphi$ is the polarisation angle of the drive and $P_{n}$ is the amplitude of its $n$-th harmonic; $n=0$, $n=1$ and $n=3$ are the dc component, the fundamental and the third harmonic.

At $V_{z}=0$ the Hamiltonian is invariant under $C_{2y}$, which reverses the layer polarisation, $P_{z}\to-P_{z}$, and acts on the polarisation vector of the drive as $\hat{e}_{\varphi}\to-\hat{e}_{-\varphi}$.  A global sign of a monochromatic field is a half-period shift, $\mathbf{E}(t+T/2)=-\mathbf{E}(t)$, so the drive at $\varphi$ is carried onto the drive at $-\varphi$ at time $t+T/2$, and
\begin{equation}
P_{z}(t;\varphi)=-P_{z}(t+T/2;-\varphi).
\label{eq:SM:parity}
\end{equation}
Inserting Eq.~(\ref{eq:SM:harmonics}) and using $e^{-in\omega(t+T/2)}=(-1)^{n}e^{-in\omega t}$ gives the selection rule for the harmonic amplitudes,
\begin{equation}
P_{n}(\varphi)=-(-1)^{n}P_{n}(-\varphi).
\label{eq:SM:evenodd}
\end{equation}
For even $n$, Eq.~(\ref{eq:SM:evenodd}) makes $P_{n}$ odd in $\varphi$, while for odd $n$ it makes $P_{n}$ even.

The $n$-th harmonic is the contraction of a $C_{3}$-invariant symmetric tensor of rank $n$ with $n$ powers of the polarisation vector, so its angular dependence is restricted by $C_{3z}$ and, at low order, is isotropic: the rank-two tensor is a multiple of $\mathbf{1}$ (Rule 2), and the rank-zero amplitude is a constant.  Hence $P_{0}=P_{2}=0$ by Eq.~(\ref{eq:SM:evenodd}), which removes the dc component and the second harmonic, and the fundamental is removed by Rule 1.  For $n=3$, Eq.~(\ref{eq:SM:evenodd}) makes $P_{3}$ even in $\varphi$.  The general cubic form in two dimensions is a combination of the harmonics $\cos\varphi$, $\sin\varphi$, $\cos3\varphi$ and $\sin3\varphi$; $C_{3}$ invariance eliminates the two $\varphi$-harmonics and leaves $P_{3}=a\cos3\varphi+b\sin3\varphi$, whose even part is the $\cos3\varphi$ term, so that $b=0$.  The leading averaged response to a linearly polarised drive is therefore
\begin{equation}
P_{z}^{(3\omega)}(t)\propto E_{0}^{3}\cos3\varphi\cos(3\omega t),
\label{eq:SM:thirdharmonic}
\end{equation}
in agreement with Eq.~(\ref{eq:thirdharmonic}) of the main text.  The third harmonic is the component of the layer polarisation that oscillates at three times the drive frequency, it radiates at $3\omega$, and its amplitude carries the sixfold anisotropy of the $C_{3}$-symmetric band.

\smsection{Additional numerical results}{sec:SM:numerics}

\begin{figure}[htbp]
\centering
\includegraphics[width=0.7\columnwidth]{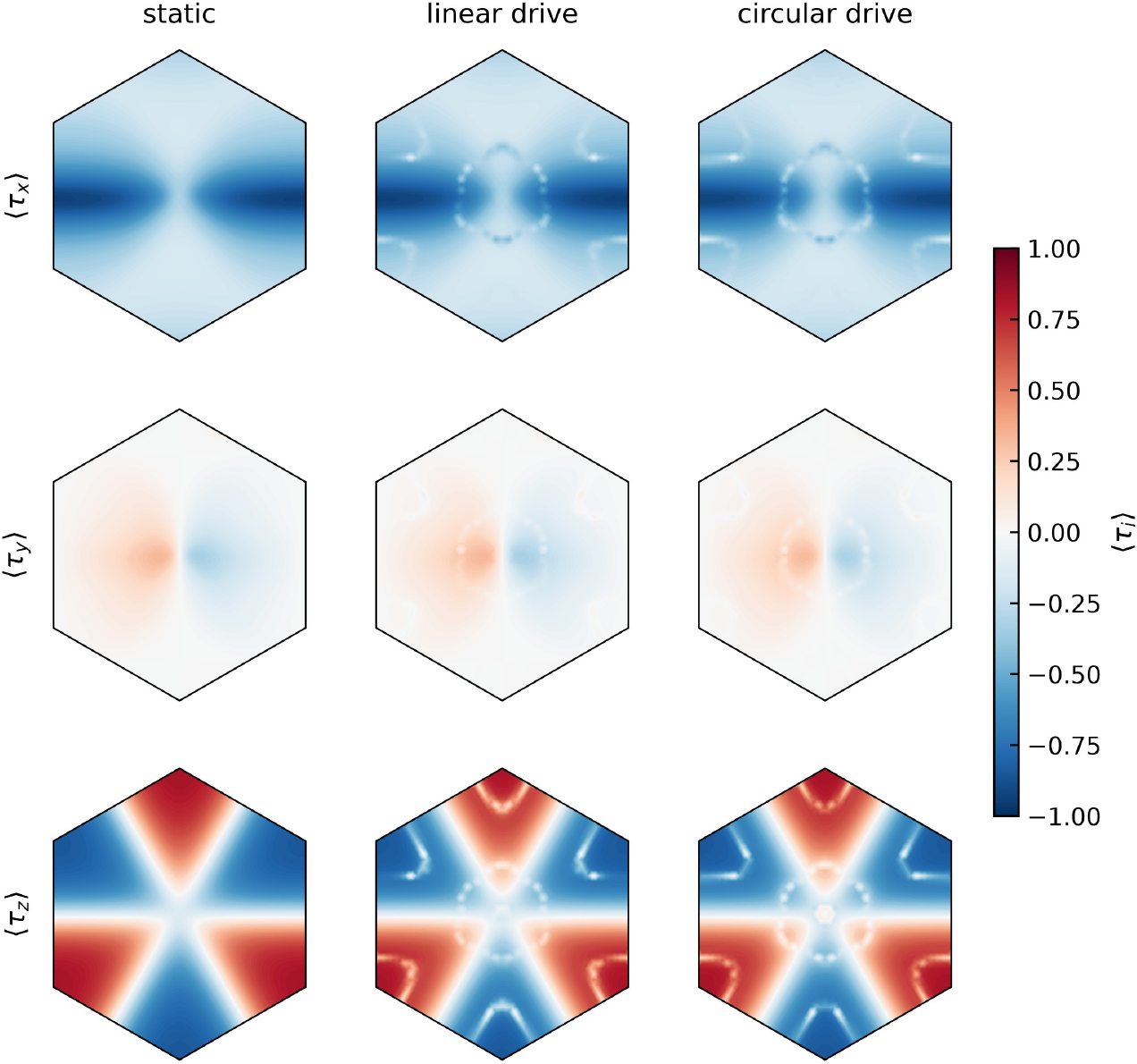}
\caption{Layer pseudo-spin on the full hexagonal mBZ, resolved into the components $\langle\tau_{x}\rangle$, $\langle\tau_{y}\rangle$ and $\langle\tau_{z}\rangle$ (rows) for the static system (left), the linear drive (middle) and the circular drive (right).  The colour scale is in units of the pseudo-spin and the panels carry no axes frame.  Parameters: $\theta=4^{\circ}$, $\hbar\omega=26$\,meV, $A_{0}=0.005$\,\AA$^{-1}$, $V_{z}=-2$\,meV.}
\label{fig:texture}
\end{figure}

\begin{figure}[htbp]
\centering
\includegraphics[width=0.7\textwidth]{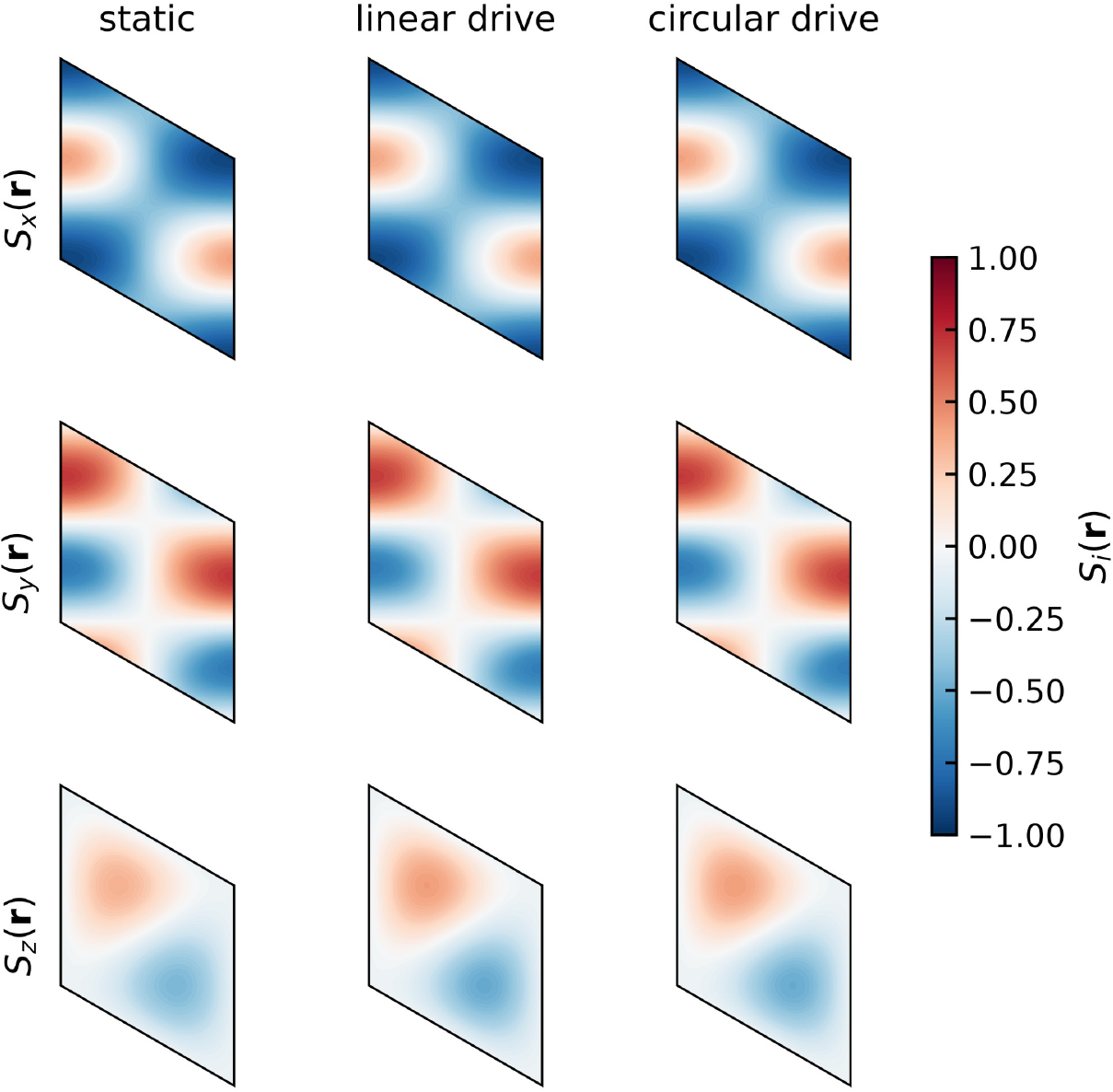}
\caption{Local layer pseudo-spin $S_{x}(\mathbf{r})$, $S_{y}(\mathbf{r})$ and
$S_{z}(\mathbf{r})$ in one moir\'{e} unit cell (rows), for the static system
(left), the linear drive (middle) and the circular drive (right).  Colour
scale in units of the layer pseudo-spin, and the panels carry no axes frame;
the drive parameters are those of Fig.~\ref{fig:texture}.}
\label{fig:SM:realspacecomponents}
\end{figure}

\end{document}